\documentclass[prl,superscriptaddress,amsmath,amssymb,twocolumn]{revtex4-2}
\usepackage{natbib}
\usepackage{graphicx}         
\usepackage{dcolumn}          
\usepackage{bm}               
\usepackage{upgreek}
\usepackage{booktabs} 				
\usepackage{tabularx}
\usepackage{array}
\usepackage{ulem}
\usepackage{float}
\usepackage[colorlinks=true,urlcolor=blue,breaklinks=true]{hyperref}
\usepackage{xcolor}
\usepackage[a4paper,left=2cm,right=2cm]{geometry}

\begin{document}
\title{Current-induced magnetization switching in a magnetic topological insulator heterostructure}

\author{Erik Zimmermann}
\thanks{These authors contributed equally to this work.}
\affiliation{Peter Gr\"unberg Institut (PGI-9), Forschungszentrum J\"ulich, 52425 J\"ulich, Germany}
\affiliation{JARA-Fundamentals of Future Information Technology, J\"ulich-Aachen Research Alliance, Forschungszentrum J\"ulich and RWTH Aachen University, Germany}

\author{Justus Teller}
\thanks{These authors contributed equally to this work.}
\affiliation{Peter Gr\"unberg Institut (PGI-9), Forschungszentrum J\"ulich, 52425 J\"ulich, Germany}
\affiliation{JARA-Fundamentals of Future Information Technology, J\"ulich-Aachen Research Alliance, Forschungszentrum J\"ulich and RWTH Aachen University, Germany}

\author{Michael Schleenvoigt}
\affiliation{Peter Gr\"unberg Institut (PGI-9), Forschungszentrum J\"ulich, 52425 J\"ulich, Germany}
\affiliation{JARA-Fundamentals of Future Information Technology, J\"ulich-Aachen Research Alliance, Forschungszentrum J\"ulich and RWTH Aachen University, Germany}

\author{Gerrit Behner}
\affiliation{Peter Gr\"unberg Institut (PGI-9), Forschungszentrum J\"ulich, 52425 J\"ulich, Germany}
\affiliation{JARA-Fundamentals of Future Information Technology, J\"ulich-Aachen Research Alliance, Forschungszentrum J\"ulich and RWTH Aachen University, Germany}

\author{Peter Sch\"uffelgen}
\affiliation{Peter Gr\"unberg Institut (PGI-9), Forschungszentrum J\"ulich, 52425 J\"ulich, Germany}
\affiliation{JARA-Fundamentals of Future Information Technology, J\"ulich-Aachen Research Alliance, Forschungszentrum J\"ulich and RWTH Aachen University, Germany}

\author{Hans L\"uth}
\affiliation{Peter Gr\"unberg Institut (PGI-9), Forschungszentrum J\"ulich, 52425 J\"ulich, Germany}
\affiliation{JARA-Fundamentals of Future Information Technology, J\"ulich-Aachen Research Alliance, Forschungszentrum J\"ulich and RWTH Aachen University, Germany}

\author{Detlev Gr\"utzmacher}
\affiliation{Peter Gr\"unberg Institut (PGI-9), Forschungszentrum J\"ulich, 52425 J\"ulich, Germany}
\affiliation{JARA-Fundamentals of Future Information Technology, J\"ulich-Aachen Research Alliance, Forschungszentrum J\"ulich and RWTH Aachen University, Germany}

\author{Thomas Sch\"apers}
\email{th.schaepers@fz-juelich.de}
\affiliation{Peter Gr\"unberg Institut (PGI-9), Forschungszentrum J\"ulich, 52425 J\"ulich, Germany}
\affiliation{JARA-Fundamentals of Future Information Technology, J\"ulich-Aachen Research Alliance, Forschungszentrum J\"ulich and RWTH Aachen University, Germany}
\hyphenation{}
\date{\today}

\begin{abstract}
We present the current-induced switching of the internal magnetization direction in a magnetic topological insulator/topological insulator heterostructure in the quantum anomalous Hall regime. The switching process is based on the bias current dependence of the coercive field, which is attributed to the effect of the spin-orbit torque provided by the unpolarized bias current. Increasing the bias current leads to a decrease in the magnetic order in the sample. When the applied current is subsequently reduced, the magnetic moments align with an externally applied magnetic field, resulting in repolarization in the opposite direction. This includes a reversal of the spin polarisation and hence a reversal of the chiral edge mode. Possible applications in spintronic devices are discussed. 
\end{abstract}
\maketitle

\section{Introduction}\setlength{\parskip}{0pt}
The quantum anomalous Hall effect (QAHE) is characterized by a vanishing longitudinal resistivity and a quantization of the Hall resistance at $h/e^2\approx 25.8\,$k$\Omega$~\cite{chang2023colloquium}. In this regime a single chiral edge channel and a bulk-insulating behaviour are created by the interplay of electrons and an internal magnetization pointing perpendicular to the sample plane~\cite{oh2013complete}. This effect was first observed in 2013 by Chang \textit{et al.}~\cite{chang2013experimental} in thin films of ferromagnetic Cr-doped topological insulator material (Bi,Sb)$_2$Te$_3$. Since then, the QAHE has been observed at low temperatures in several Cr- and V-based magnetic topological insulators (MTIs)~\mbox{\cite{chang2023colloquium,zhang2017ferromagnetism,chang2015high,lippertz2022current,mogi2015magnetic,kawamura2020current}}. The related chiral edge mode could possibly be a suitable basis for Majorana-based applications when combined with a superconductor, which makes MTIs promising candidates for future topological quantum computation operations~\cite{adagideli2020time,hassler2020half,liu2018majorana,chen2018quasi,zeng2018quantum,beenakker2019deterministic}.

The relevance of the applied current in MTI-based structures is underlined by recent reports about the break down of the QAHE dependent on the applied current and the dimensions of the sample. The latter gains importance, especially when aiming for nanostructures~\cite{lippertz2022current,fox2018part}. In this work we investigate the current dependence of MTI magnetotransport properties. Current sweeps at fixed magnetic fields are used to switch the branch in the hysteresis only by changing the bias current. The experiment is completed by measurements at several different magnetic fields which allow a detailed investigation of the switching properties. The underlying mechanism of the current-induced magnetization switching is attributed to the current dependence of the coercive magnetic field $\mu_0H_c$.

\section{Experimental}
\begin{figure}[hbtp]
\centering
\includegraphics[width=0.47\textwidth]{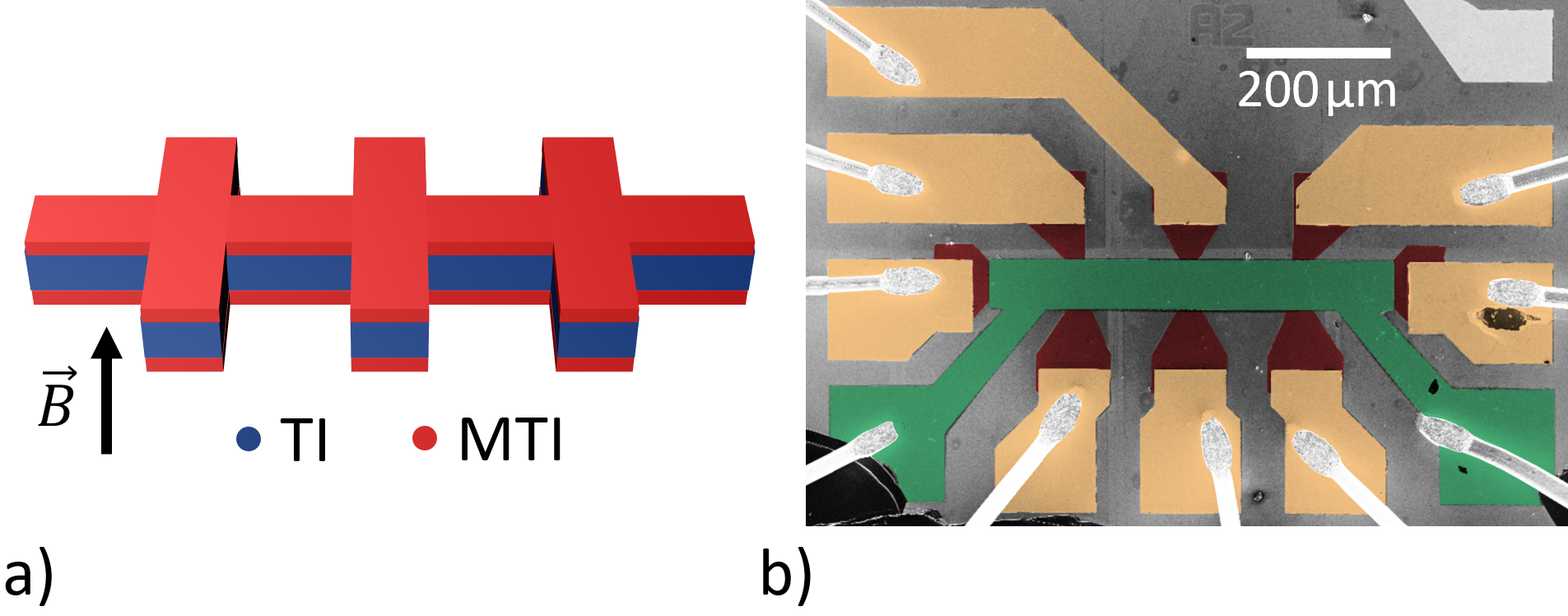}
\caption{MTI Hall bar sample. a) A sketch of the Hall bar indicates the layering of MTI and conventional TI in the heterostructure. b) A false-color scanning electron microscope image of the Hall bar depicts the MTI Hall bar (red), the top gate (green) and the metal contacts (orange). Note that the center part of the Hall bar is completely covered by the gate and has a width of $10\,\upmu$m.}\label{Fig_Sandwich_Sample}
\end{figure}

\setlength{\parskip}{0pt}
As magnetic topological insulator material Cr$_{0.21}$Bi$_{0.51}$Sb$_{1.28}$Te$_3$ is used. In order to reduce possible contributions from a bulk channel caused by additional disorder arising from the chromium, a tri-layer heterostructure is formed~\cite{mogi2015magnetic}. As shown in the sketch of the Hall bar structure in Fig.~\ref{Fig_Sandwich_Sample} a), a non-magnetic topological insulator (TI) layer is sandwiched in-between a top and bottom MTI layer to form an MTI/TI heterostructure. Here, a ternary compound with a stoichiometry of Bi$_{0.55}$Sb$_{1.45}$Te$_3$ is chosen for the non-magnetic TI interlayer.

For the growth of the MTI/TI heterostructure, a Si(111) substrate is cleaned using a Piranha solution followed by $1\,$\% hydrofluoric acid in order to remove native oxides and passivate the surface. Subsequently, the sample is loaded into a molecular beam epitaxy (MBE) chamber and heated to $700\,^\circ$C to desorb the hydrogen passivation. After cooling the substrate down to the growth temperature of $220\,^\circ$C within $30$ minutes, first a Te flux is supplied for $30\,$s to re-passivate the surface and prepare the sample for the subsequent layer growth. Next, Cr is provided from a standard Knudsen cell, set to $1045\,^\circ$C, while Bi and Sb fluxes are added in a ratio of about Bi:Sb:Te 1:3:42 resulting in a MTI growth rate of $0.33\,$nm/min. Rutherford backscattering spectrometry (RBS) measurements on MTI films grown with these parameters yield the aforementioned stoichiometry of Cr$_{0.21}$Bi$_{0.51}$Sb$_{1.28}$Te$_3$. Additionally, layers grown with the same parameters but without Cr flux yield a stoichiometry of Bi$_{0.55}$Sb$_{1.45}$Te$_3$ via RBS. The tri-layer heterostructure is created by supplying Bi and Sb for $1\,$min to initialize TI growth, adding Cr flux for $5\,$min, closing the Cr shutter for $12\,$min and lastly reopening the shutter for another $5\,$min. The total thickness of the tri-layer amounts to $7.3\,$nm, divided into $1.6\,$nm MTI/$4\,$nm TI/$1.6\,$nm MTI, nominally. As a last step, a layer of $5\,$nm Al$_2$O$_3$ is deposited in-situ via electron beam evaporation in order to prevent the sample from oxidation.

Subsequently, the plain film is etched into a Hall bar shape using reactive ion etching and $50\,$nm Ti/$100\,$nm Pt metal contacts are deposited via metal sputtering. The Hall bar structure has a width of $10\,\upmu$m. The neighbouring side contacts of the Hall bar are spaced by a distance of $150\,\upmu$m. Using atomic layer deposition $15\,$nm of HfO$_2$ are deposited globally as a dielectric layer. Lastly, a $50\,$nm Ti/$100\,$nm Au gate is placed on top. The final device is depicted in Fig.~\ref{Fig_Sandwich_Sample} b).

The magnetotransport measurements are performed in a variable temperature insert cryostat with a base temperature of $1.3\,$K and a $14\,$T superconducting magnet. The magnetic field $\mu_0 H$ is oriented perpendicular to the substrate. All measurements are conducted using standard lock-in techniques in a four-terminal setup at base temperature. A voltage of $2\,$V is applied to the top gate for all measurements since under these conditions the sample is in the regime of the QAHE (cf. supplementary material).

\section{Results and Discussion}

\subsection{Current dependence}
\begin{figure*}[hbtp]
\centering
\includegraphics[width=0.94\textwidth]{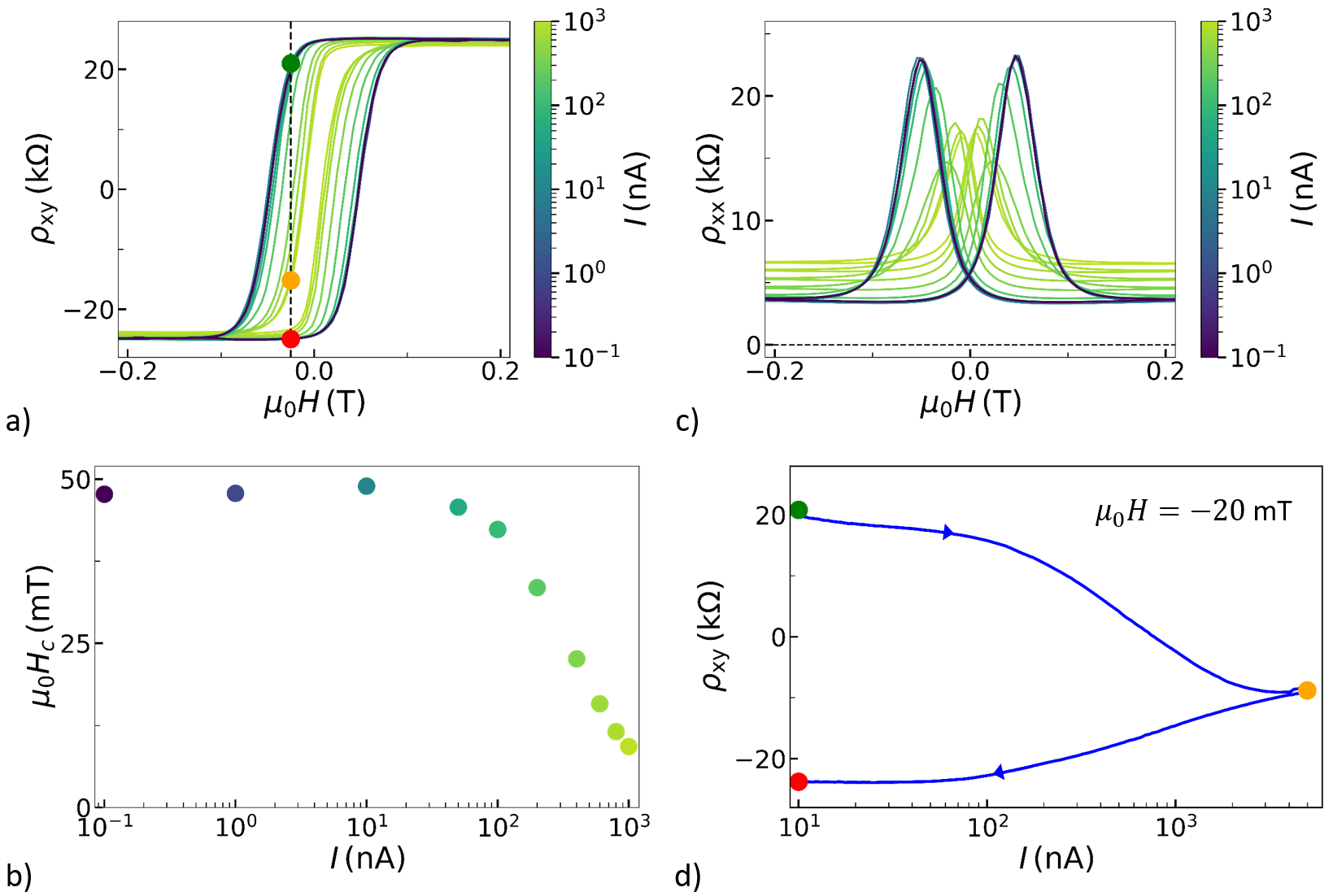}
\caption{Current dependence of the magnetotransport properties of the MTI/TI heterostructure. a) Transversal resistivity $\rho_{xy}$ as a function of magnetic field for different applied currents: 0.1, 1.0, 10, 50, 100, 200, 400, 600, 800, and 1000\,nA. Note: The color code for the current is logarithmic. b) The current dependence of the coercive magnetic field $\mu_0H_c$ is plotted. c) Corresponding longitudinal resistivity $\rho_{xx}$ as a function of the magnetic field for the same current values as in a). d) For a constant magnetic field of $-20\,$mT starting at the green dot the current is ramped up from 10\,nA to $5\,\upmu$A (yellow dot). Afterwards, the current is lowered again to the initial value of 10\,nA (red dot).}\label{Fig_Sandwich_I_dep}
\end{figure*}

First, the current dependence of the magnetotransport properties of the Hall bar sample is presented. Figure~\ref{Fig_Sandwich_I_dep} a) shows the transversal resistivity $\rho_{xy}$ as a function of a perpendicular magnetic field for different applied currents ranging non-linearly from $100\,$pA up to $1\,\upmu$A. All curves show a pronounced hysteresis. The coercive magnetic field $\mu_0H_c$ substantially decreases with increasing current above 10\,nA, indicating a weakening of the magnetism in the sample. The current dependence of $\mu_0H_c$ is shown in Fig.~\ref{Fig_Sandwich_I_dep} b). The transversal resistivity saturates at a value of 25\,k$\Omega$ at a bias current of 0.1\,nA close to the quantized value of $h/e^2$. Increasing the bias current results in a gradual decrease of $\rho_{xy}$. Similar results are presented by Kawamura \textit{et al.}~\cite{kawamura2020current}. Since we found a saturation value $\rho_{xy}$ close to $h/e^2$, transport can be assumed to take place in the QAHE regime with a high current density in the chiral edge channel. However, due to the finite longitudinal resistivity a small current contribution can be assigned to bulk transport.  

Figure~\ref{Fig_Sandwich_I_dep}~c) shows the longitudinal resistivity $\rho_{xx}$ for different applied currents. The longitudinal resistivity also shows a hysteretic behavior with a peak at the magnetic field position where the magnetization reverses. As indicated by the colors in the plot, one can see that the shape of the hysteresis changes mostly for currents above $10\,$nA. With increasing bias current the peak that indicates the switching of the magnetization direction becomes less pronounced and the remaining resistivity at elevated magnetic fields is further increased. The finite resistance, as well as the increase in resistance with increasing applied current at elevated magnetic fields can be explained by the presence of local charge puddles in-between the edge channels on both sides of the Hall bar structure~\cite{ito2022cancellation,knispel2017charge,borgwardt2016self}. Here, the remaining thermal energy and the applied potential arising from the current and the finite resistance cause a growth of the puddles leading to a conducting connection across the sample and therefore create a finite resistance~\cite{lippertz2022current}.

Two possible origins of the dependence of the coercive magnetic field on the current are suggested by Khang \textit{et al.}~\cite{khang2018conductive}: On the one hand, Joule heating arising from the previously mentioned enhanced current density could cause a decrease in the binding energy of the magnetic moments and thus a decrease of $H_c$. On the other hand, a spin–orbit torque caused by spin-injection could be a possible reason. Furthermore, Lippertz \textit{et al.}~\cite{lippertz2022current} concluded a break down of the QAHE because of the presence of two-dimensional charge puddles that combine to a conducting channel at higher biases. In our case, Joule heating due to an increased applied current in the chiral edge mode is excluded, as this would also lead to a significant break down of the Hall resistivity at elevated magnetic fields. Although we expect the presence of local charge puddles in our material, we exclude them as well as a dominant origin for the break down mechanism for the same reason. Instead, when applying an unpolarized current, the spin-orbit torque can lead to a randomization of the spin-direction in the magnetic area for increased biases~\cite{marrows2005spin,khang2018conductive}. Indeed, an unpolarized current applied to a spin-polarized domain can even create a spin-polarization in opposite direction close to the magnetic domain due to spin-dependent reflection and transmission properties~\cite{dugaev2003reflection}. Another possibility is the effect of hot energetic electrons in the confined surface states that can cause a local breakdown of the spin-polarization.

\subsection{Magnetization switching}
In the previous section the impact of the applied current on the coercive field of our MTI/TI heterostructure was discussed. Until here, only magnetic field sweeps at different constant currents have been performed. Now, at constant magnetic fields the current is changed. It is expected, that the bias current affects the magnetization in the transition region around $\mu_0 H=0$ and thus the corresponding transversal resistivity $\rho_{xy}$. Three colored points at a constant magnetic field in Fig.~\ref{Fig_Sandwich_I_dep} a) indicate hallmarks of the experiment. For initialization, the magnetic moments in the sample are first polarized by a positive magnetic field so that $\rho_{xy}$ saturates. Subsequently, the magnetic field is ramped to a value of $-20\,$mT so that at small bias currents $\rho_{xy}$ settles at a value indicated by the green dot. As discussed in the previous section, an increased current lowers the coercive magnetic field. Following the dashed line in Fig.~\ref{Fig_Sandwich_I_dep}~a), $\rho_{xy}$ decreases with increasing applied current. If the current is sufficiently large, the resistivity decreases from positive to negative values (yellow dot). This corresponds to a reversal of the majority of the magnetic momenta in our MTI/TI heterostructure. When finally lowering the current to its initial value, more magnetic momenta in the MTI get aligned to the externally applied magnetic field. The result is a complete reversal of magnetization (red dot).

An exemplarily sequence of changing $\rho_{xy}$ by driving the current back and forth between low and high values can be traced in Fig.~\ref{Fig_Sandwich_I_dep} d). The sequence starts with a current of $I=10\,$nA at a constant magnetic field of $\mu_0 H=-20\,$mT (green dot). Due to the curvature of the hysteresis the value of $\rho_\text{xy}$ has already reduced slightly compared to its saturation value. As the current is increased, one finds that first the applied current slowly reduces the value of $\rho_\text{xy}$. When passing $\rho_\text{xy}=0$ the current reduces the coercive magnetic field below the absolute value of the externally applied magnetic field. Upon increasing the current to $5\,\upmu$A the magnetic momenta predominantly switch their direction and align with the external magnetic field (yellow dot). When reducing the current again, the polarization is fully reversed (red dot). Thus, switching the direction of the magnetization is achieved by only varying the applied current. Hereby, the result for the transport is that the chiral edge mode is inverted and thus also its spin polarization.

\begin{figure*}[hbtp]
\centering
\includegraphics[width=0.94\textwidth]{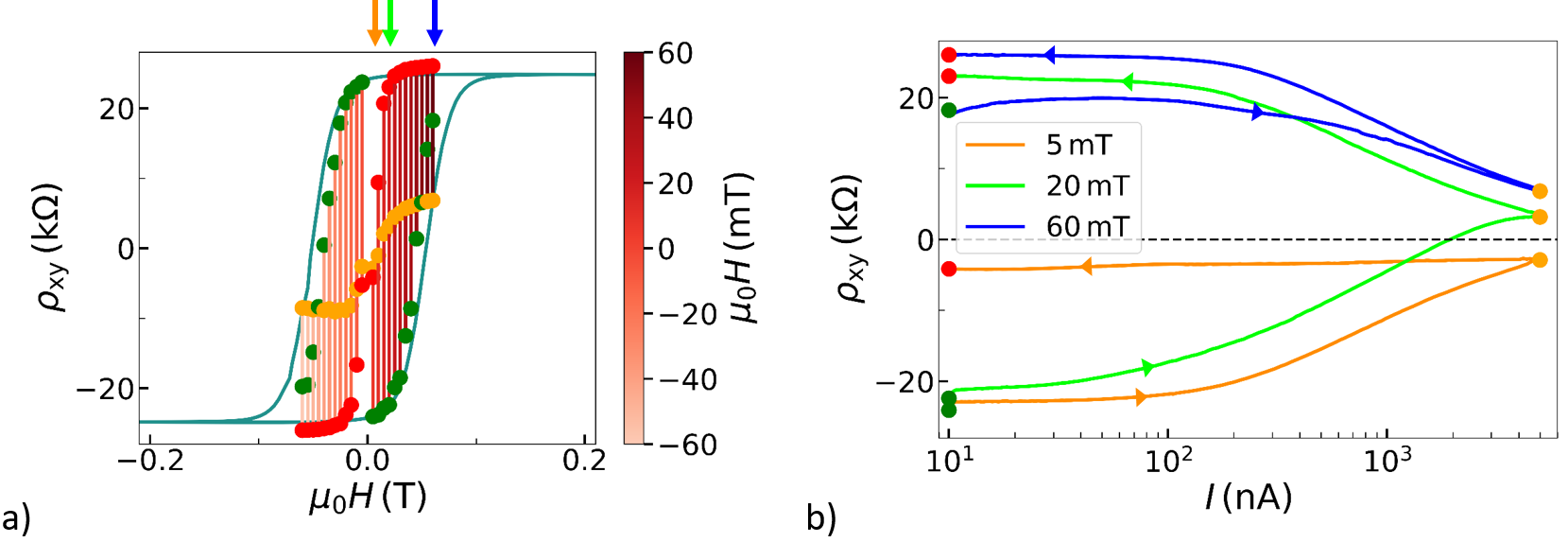}
\caption{Current-induced magnetization switching experiment for different magnetic bias fields. a) The curves show $\rho_{xy}$ for current sweeps from $10\,$nA (green dots) to $5\,\upmu$A (yellow dots) back to $10\,$nA (red dots) at constant magnetic fields ranging from $-60\,$mT to $60\,$mT in steps of $5\,$mT. The hysteresis of $\rho_{xy}$ for a bias current of $10\,$nA, taken from Fig.~\ref{Fig_Sandwich_I_dep}~a), is plotted as a reference. b) The current dependence of three representative curves from a) at 5, 20, and $60\,$mT are compared.}\label{Fig_Sandwich_CIMS_exp}
\end{figure*}

In order to get a deeper insight into the switching mechanism, the sequence described in connection with Fig.~\ref{Fig_Sandwich_I_dep} d) is run for different fixed magnetic fields $|\mu_0 H|\leq 60\,$mT with a stepping of $5\,$mT. In Fig.~\ref{Fig_Sandwich_CIMS_exp} a) the initial and final values of $\rho_{xy}$ (red and green dots, respectively) as well as the values at the maximum applied bias current (yellow dots) are plotted at the according magnetic field values, similarly to the ones in Fig.~\ref{Fig_Sandwich_I_dep} a). As a reference the $10\,$nA hysteresis curve from Fig.~\ref{Fig_Sandwich_I_dep}~a) is also plotted in the same graph. The green dots representing the starting values lie quite well on the reference hysteresis curve, as expected. Whereas, the curve following the red dots for the final values is centered around zero magnetic field. Compared to that, the curve following the yellow dots corresponding to the maximum bias current of 5\,$\mu$A comprises a smaller amplitude swing. One finds that for magnetic fields not too close to zero, i.e. larger than about 10\,mT, and smaller than the coercive field a switching occurs, since at elevated currents the sign of $\rho_{xy}$ and by that the direction of the magnetization is inverted. Upon, returning the bias current to 10\,nA this magnetization orientation is kept. Hence, there is a successful switching of the branch obtained only by applying a pulse of a larger current.

In Fig.~\ref{Fig_Sandwich_CIMS_exp} b) the data for three representative curves is plotted  against the applied current. Starting in the lower branch of the hysteresis, for curves at moderate magnetic fields at 20\,mT, the same switching behaviour as in Fig.~\ref{Fig_Sandwich_I_dep} d) is found, i.e. the sign of $\rho_{xy}$ and the direction of magnetization is inverted. In contrast, very close to zero magnetic field it is observed that the switching process fails, as indicated in Fig.~\ref{Fig_Sandwich_CIMS_exp}~b) exemplarily by the curve recorded at $\mu_0 H=5\,$mT. Thus, at small magnetic fields the external magnetic field is insufficient to realign the magnetic moments in a ferromagnetic order. Also for fields larger than the coercive field, i.e. 60\,mT in Fig.~\ref{Fig_Sandwich_CIMS_exp} b), where the magnetization is already aligned to the external field, the magnetization is weakened by the bias current but keeps its orientation. After returning to 10\,nA even the saturation value of $\rho_{xy}$, is reached, which is larger that the initial value.

It becomes apparent that the increased current weakens or even destroys the alignment of the magnetic moments, as indicated by the sequence of yellow dots in Fig.~\ref{Fig_Sandwich_CIMS_exp} a) being lower in amplitude than the initial hysteresis curve. When the current is lowered again the external magnetic field is responsible for the realignment in opposite direction. For the curves at low and high magnetic fields one can see that no switching across zero occurs, but $|\rho_{xy}|$ first decreases below the starting value due to a reduction on magnetic order caused by the the applied current for a given external magnetic field. Nonetheless, using this procedure also for the curves taken at larger magnetic fields, a switching of the other branch of the hysteresis occurs and thus stabilizing the magnetization to the saturation value. In this sense, the current bias cannot only be employed to reverse the magnetization but in general be used to switch the magnetization to saturation with a significantly reduced external magnetic field.   

\section{Conclusion} \setlength{\parskip}{0pt}
The current dependence of the coercive magnetic field of an MTI/TI heterostructure was employed in order to switch the direction of the internal magnetization by varying the bias current. It is concluded that after the reduction of the magnetic order with the use of an applied current the magnetic moments select the energetically favorable state which corresponds to an alignment along the external magnetic field.
It is concluded that no magnetization is created in the process. Instead, only the direction is switched when comparing start and end point. Around zero magnetic field, where the procedure fails, it can be seen that the effect is caused by an interplay of current ramp and external magnetic field as here, the external magnetic field is insufficient to re-polarize the magnetic momenta. For magnetic fields above the coercive fields no reversal of the magnetization takes place. However, in this case a stabilization into the other branch of the hysteresis is achieved. 

In other current-induced magnetization switching experiments the magnetization direction can be switched back to the initial orientation by reversing the current. For the presented sample, however, the switching back to the initial magnetisation needs a reversal of the externally applied magnetic field~\cite{cai2017electric,liu2019current}. Nevertheless, the switching of the direction of the spin-polarized edge channel may also have interesting applications in spintronics. One apparent advantage is that current-induced magnetic switching at a fixed external magnetic field is technically easier realized than switching by variation of a magnetic field. Apart from the bias current induced reversal of the magnetization in a certain magnetic field range, in general the current bias can also be employed to switch the magnetization to a stable state at significantly reduced external magnetic fields. It is indeed remarkable that without the assistance of the current sweep much larger magnetic fields are necessary to switch to the other branch of the $\rho_{xy}$ hysteresis. This current-induced switching mechanism might be interesting for magnetic memory applications. In future experiments the switching effect observed here should be investigated with samples of different sizes and geometries in order to create tailored spintronic devices.

\section{Acknowledgments}
We thank Herbert Kertz for technical assistance. All samples have been prepared at the Helmholtz Nano Facility~\cite{albrecht2017hnf}. This work is funded by the Deutsche Forschungsgemeinschaft (DFG, German Research Foundation) under Germany's Excellence Strategy – Cluster of Excellence Matter and Light for Quantum Computing (ML4Q) EXC 2004/1 – 390534769, by the German Federal Ministry of Education and Research (BMBF) via the Quantum Futur project ‘MajoranaChips’ (Grant No. 13N15264) within the funding program Photonic Research Germany and by the QuantERA grant MAGMA via the German Research Foundation under grant 491798118.
\noindent 
%

\clearpage
\widetext

\setcounter{section}{0}
\setcounter{equation}{0}
\setcounter{figure}{0}
\setcounter{table}{0}
\setcounter{page}{1}
\makeatletter
\renewcommand{\thesection}{S\Roman{section}}
\renewcommand{\thesubsection}{\Alph{subsection}}
\renewcommand{\theequation}{S\arabic{equation}}
\renewcommand{\thefigure}{S\arabic{figure}}
\renewcommand{\figurename}{Supplementary Figure}
\renewcommand{\bibnumfmt}[1]{[S#1]}
\renewcommand{\citenumfont}[1]{S#1}

\begin{center}
\textbf{Current-induced magnetization switching in a magnetic topological insulator heterostructure\\(Supplementary Material)}
\end{center}

\section{Temperature dependence and normal transport of the MTI Hall bar}
\begin{figure}[]
\centering
\includegraphics[width=\textwidth]{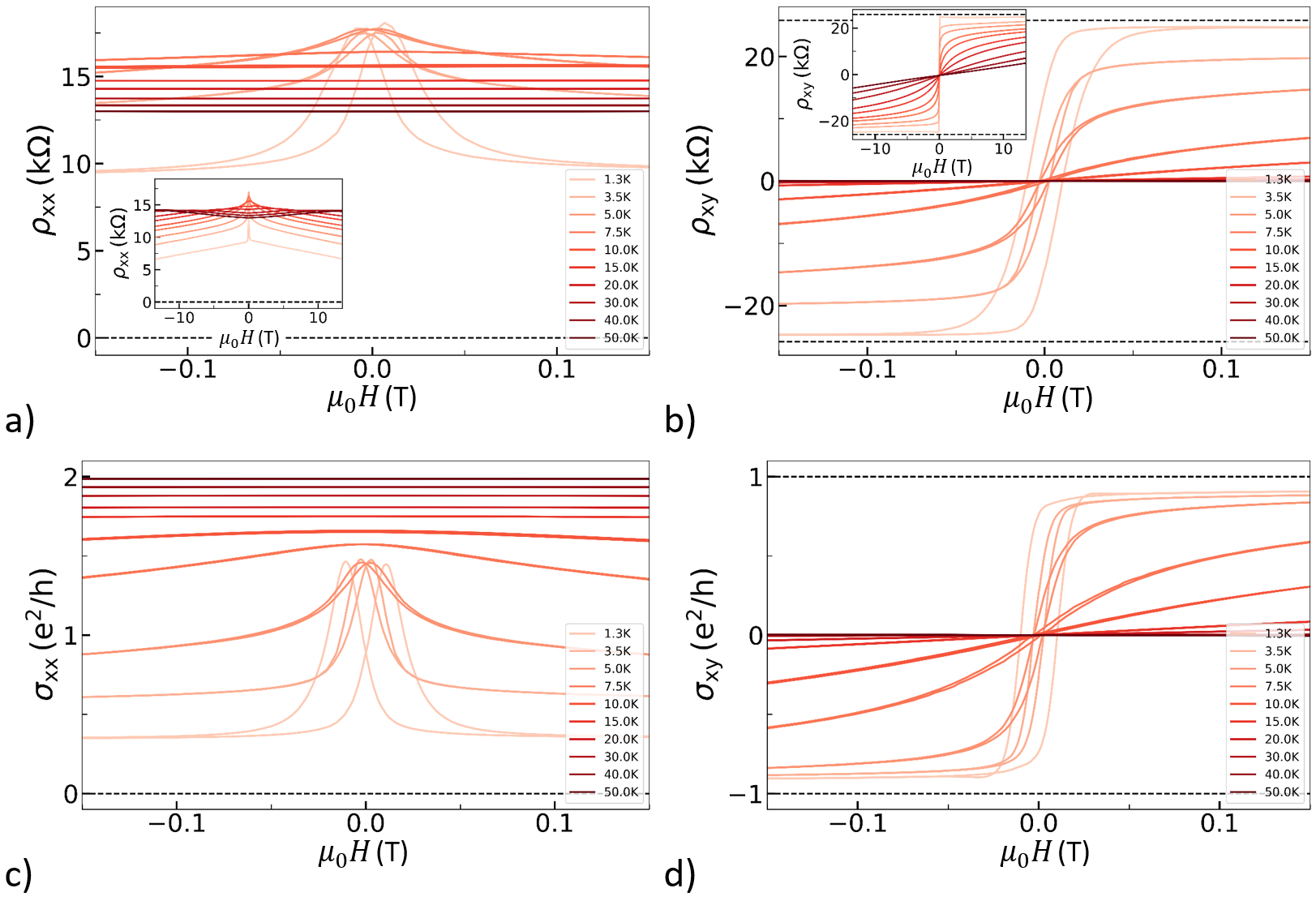}
\caption[Temperature dependent data of the MTI heterostructure]{Temperature dependent data of the MTI heterostructure. a) The temperature dependence of the longitudinal resistivity around zero magnetic field is shown with the inset covering a larger magnetic field range. b) The corresponding Hall data shows values close to $25.8\,$k$\Omega$ (dashed lines) for base temperature and the long-range behaviour is shown in the inset. In c) and d) the calculated temperature dependent longitudinal and transversal conductivity are shown, respectively.}\label{Fig_Sandwich_T_dep}
\end{figure}

Figure~\ref{Fig_Sandwich_T_dep} shows the temperature dependent magnetotransport data obtained from the magnetic topological insulator (MTI) heterostructure Hall bar. In a) the longitudinal resistivity calculated from the resistance and the dimensions of the sample is plotted. It is observed that the resistance at elevated magnetic fields of the curve at $T=1.3\,$K is significantly reduced compared to the curves at higher temperatures. In Figure~\ref{Fig_Sandwich_T_dep} b) the Hall resistivity is plotted. One can see that for base temperature the height of the hysteresis is $\rho_\text{AH}=24.9\,$k$\Omega$ which is close to the desired value of about $25.8\,$k$\Omega$. The explanation is that the spin polarized edge mode has some minor remaining losses through the bulk so that the two sides of the sample are not yet independent from each other. This is supported by the non-vanishing longitudinal resistivity. On the one hand, for base temperature the longitudinal resistivity drops below the transversal one and thus indicates a step towards the quantum anomalous Hall effect (QAHE)~\cite{chang2013experimental}. On the other hand, due to the lossless transport mediated by the chiral edge mode in the case of the QAHE a drop to zero is expected.

The inset of Figure~\ref{Fig_Sandwich_T_dep}~b) shows the long-range behaviour of the transversal raw data. It is seen that for larger magnetic fields the value for base temperature stays nearly constant while a non-linear increasing behaviour is observed for higher temperatures. From the long-range data at base temperature a charge carrier concentration of $ n_{\text{2D}}=3.2\cdot 10^{12}\,$cm$^{-2}$ and a mobility of $\mu=114.5\,$cm$^2$/Vs are derived. For comparison, using $\sigma_{xx}=\frac{\rho_{xx}}{\rho_{xx}^2+\rho_{xy}^2}$ and $\sigma_{xy}=\frac{-\rho_{xy}}{\rho_{xx}^2+\rho_{xy}^2}$ the data is converted to the longitudinal and transversal conductivity and plotted in Figure~\ref{Fig_Sandwich_T_dep}~c) and d), respectively. Here, the temperature development towards the QAHE can be seen more clearly. 

\section{Temperature dependent magnetotransport of a chromium-doped topological insulator mono-layer}
\begin{figure}[hbtp]
\centering
\includegraphics[width=\textwidth]{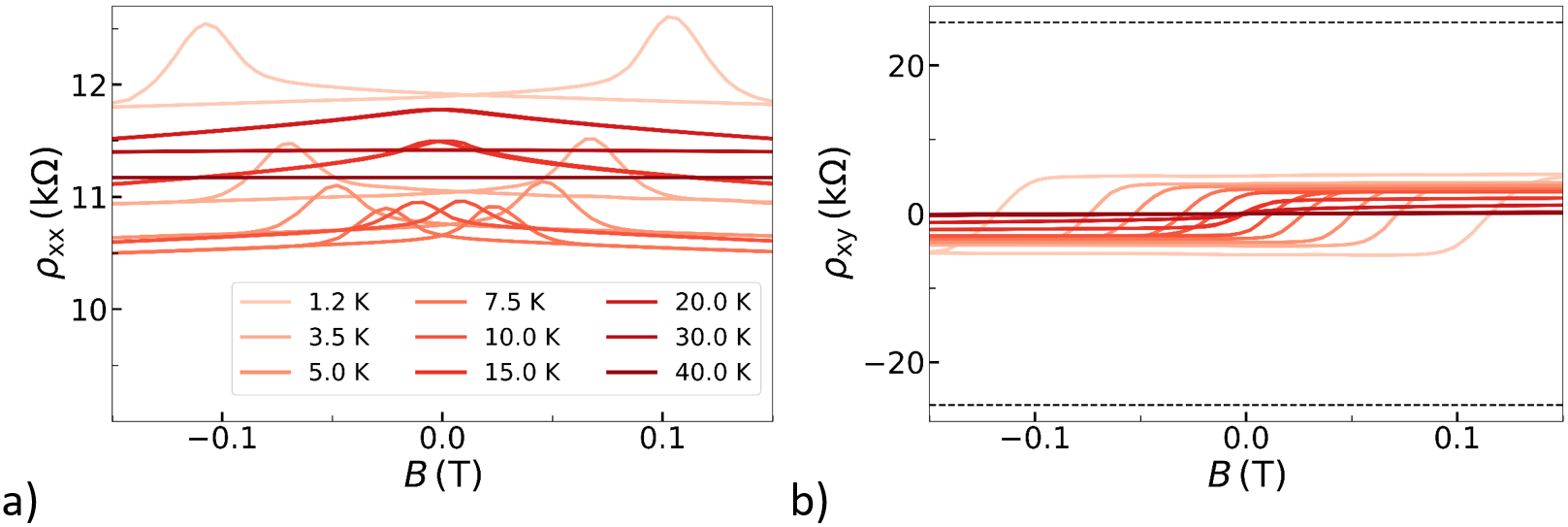}
\caption{Temperature dependent transport data of a 6\;nm thick Cr$_{0.21}$Bi$_{0.51}$Sb$_{1.28}$Te$_3$ mono-layer. a) The temperature dependence of the longitudinal resistivity shows peaks at lower temperatures which vanish up to 40\;K. All resistivity values are not close to zero indicating the not quantized anomalous Hall effect rather than the QAHE. b) The corresponding Hall data show hystereses according to the anomalous Hall effect. The saturation levels are far below 10\;k$\Omega$, thus, clearly indicating no QAHE. The coercitivity declines for elevated temperatures.}\label{Fig_mono-layer_T_dep}
\end{figure}

For the investigations in the main article a tri-layer heterostructure is used, formed out of two layers of the MTI Cr$_{0.21}$Bi$_{0.51}$Sb$_{1.28}$Te$_3$ and one layer of the non-magnetic topological insulator (TI) Bi$_{0.55}$Sb$_{1.45}$Te$_3$ in-between. Magnetic heterostructures containing additional layers of conventional TI are expected to show a more stable QAHE compared to samples formed only out of one MTI material~\cite{mogi2015magnetic}. The advantage of the heterostructure is that the electrons are free from impurities induced by magnetic dopands when being transported through the plain topological insulator layer. At the same time, they couple to the magnetism in the MTI layer via exchange interaction~\cite{yao_topological_2021}. In Figure~\ref{Fig_mono-layer_T_dep} the temperature dependent magnetotransport data taken from a 6\;nm high Cr$_{0.21}$Bi$_{0.51}$Sb$_{1.28}$Te$_3$ reference Hall bar sample is shown. It can indeed be seen that the magnetic signals (larger longitudinal resistivity and lower hysteresis saturation) are weaker than for the heterostructure (cf. Fig.~2 in the main article). However, the magnetic coercitivity is larger. This is because the magnetic moments are spatially denser due to the absence of the undoped TI layer. As a result, the general magnetisation, and therefore also the coercive magnetic field, is larger~\cite{kools_factors_1985}. The longitudinal resistivity being not close to zero and the low saturation level of the hysteresis indicate that no QAHE is present. The magnetic signals vanish up to a temperature of $40\,$K.

\section{Gate dependence}
\begin{figure}[hbtp]
\centering
\includegraphics[width=\textwidth]{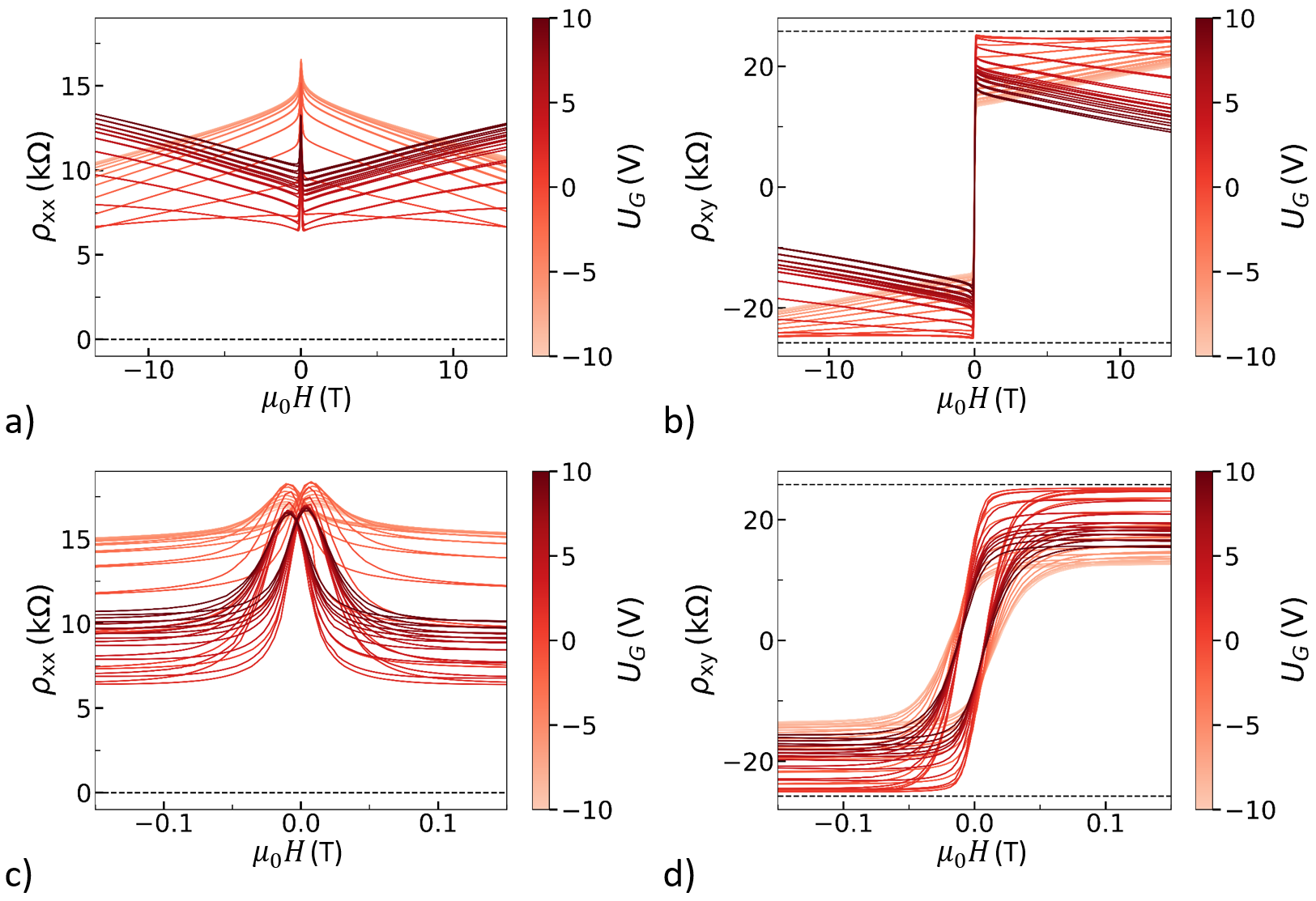}
\caption{Gate dependent data of the MTI heterostructure. In a) and b) the long-range longitudinal and transversal resistivities are plotted against the magnetic field, respectively. c) and d) depict the precise short-range measurements.}\label{Fig_Sandwich_Gate_dep}
\end{figure}

The tuning of the Fermi energy in the bulk band gap is essential in order to reduce conductance through the bulk of the sample that counteracts the QAHE~\cite{onoda2003quantized}. Experiments have confirmed that the AHE strongly depends on the position of the Fermi energy in the band structure~\cite{deng2020quantum,niu2019quantum,zhang2019experimental}. The stoichiometry of the MTI is selected so that the Fermi level is close to the Dirac point. For fine tuning, a top gate is used. Figure~\ref{Fig_Sandwich_Gate_dep} shows the gate dependent magnetotransport data recorded with an applied bias current of $I=1\,\upmu$A at a base temperature of $T=1.3\,$K. Gate voltages up to $U_G=\pm 10\,$V are applied with a stepping of $1\,$V. In Figure~\ref{Fig_Sandwich_Gate_dep} a) and b) measurements with large magnetic fields of the longitudinal and transversal resistance are shown, respectively. In c) and d) precise measurements of $\rho_\text{xx}$ and $\rho_\text{xy}$ around zero magnetic field reveal the gate dependence of the hysteresis.

Comparing the behaviour at small magnetic fields to the base temperature measurement in Figure~\ref{Fig_Sandwich_T_dep}~a), one can see that for moderate positive gate voltages the tails of the curves are lowered in resistivity. Interestingly, the long-range behaviour of the longitudinal signal changes at $U_G=2\,$V between monotonously increasing and decreasing. Due to the time reversal symmetry breaking at large magnetic fields an influence of a possible WL or WAL effect is excluded. Generally, for sufficiently large magnetic fields the resistivity of semiconductors is expected to increase approximately quadratic with increasing magnetic field~\cite{smith1978semiconductors}. This fits for high, positive gate voltages. The decreasing behaviour is explained with a dominating chiral edge mode arising from the QAHE that suppresses the longitudinal resistivity.

\begin{figure}[hbtp]
\centering
\includegraphics[width=0.7\textwidth]{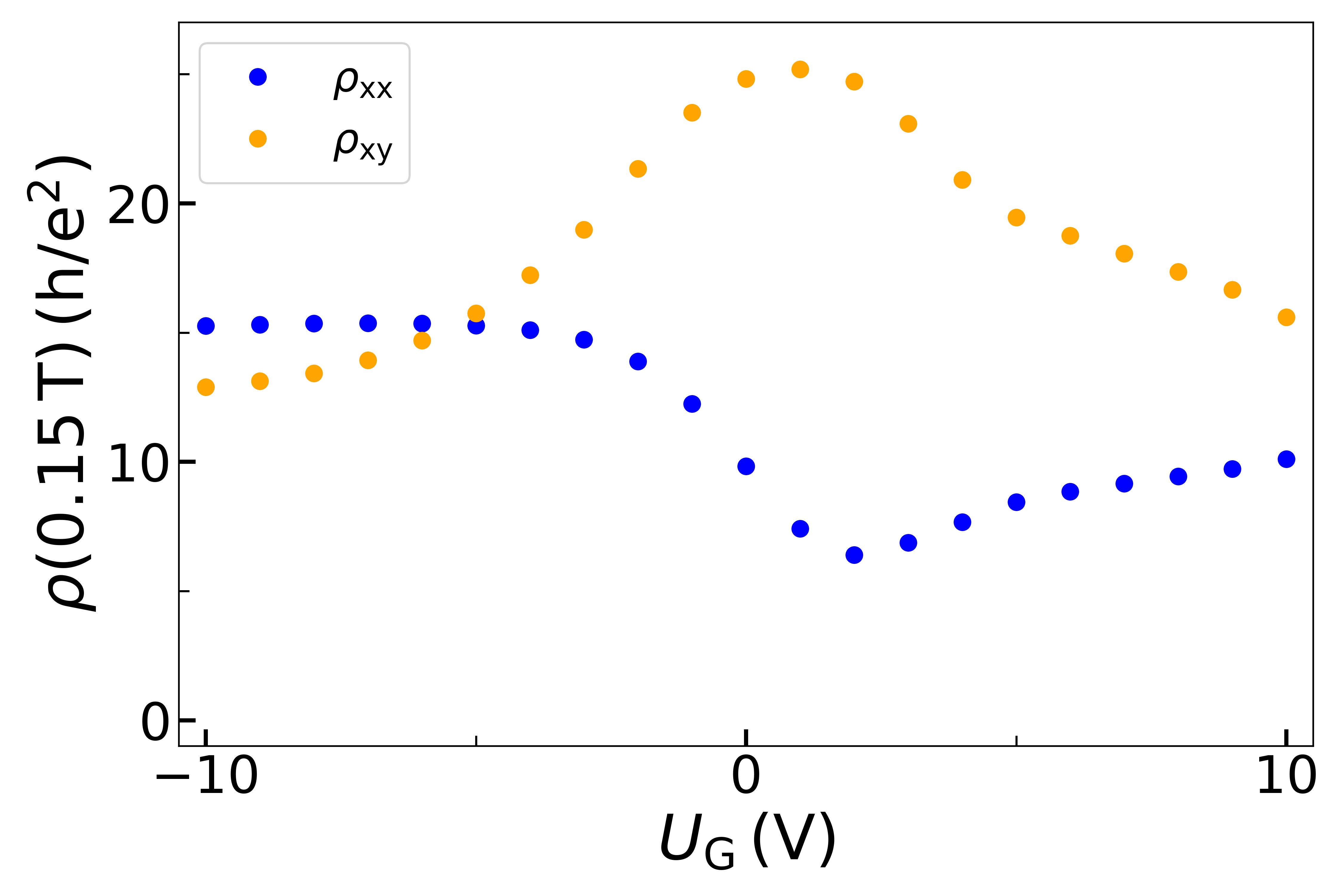}
\caption{Gate dependent longitudinal and transversal resistivity behaviour. The resistivities at $\mu_0 H=0.15\,$T are compared to each other with respect to the gate voltage.}\label{Fig_Sandwich_pxxpxy}
\end{figure}

Furthermore, it is noted that the height of the hysteresis can be changed with the gate voltage. Compared to Figure~\ref{Fig_Sandwich_T_dep}~b) a slight improvement towards quantization is observed for small positive gate voltages reaching a maximum at $U_G=1\,$V. For larger gate voltages also the Hall slope having different signs is visible and thus indicating a switch in carrier type. Together with the decrease of the hysteresis height it is an indication that the QAHE is approached best, when being closest to the Dirac point in band structure. The values of the longitudinal and transversal resistance at $\mu_0 H=0.15\,$T are compared in Figure~\ref{Fig_Sandwich_pxxpxy}. It is observed, that the maximum in Hall resistivity and the minimum in longitudinal resistivity are slightly shifted against each other.

For most of the gate voltages the value of the $\rho_\text{xy}$ is larger than the one of $\rho_\text{xx}$ and thus reproducing the results from Chang \textit{et al.}~\cite{chang2013experimental}. Still, there are some small losses indicating that even when applying a gate voltage the complete quantization is not reached, yet. For further optimization, the temperature and applied current need to be decreased.

\section{Longitudinal current dependence}
\begin{figure*}[hbtp]
\centering
\includegraphics[width=0.94\textwidth]{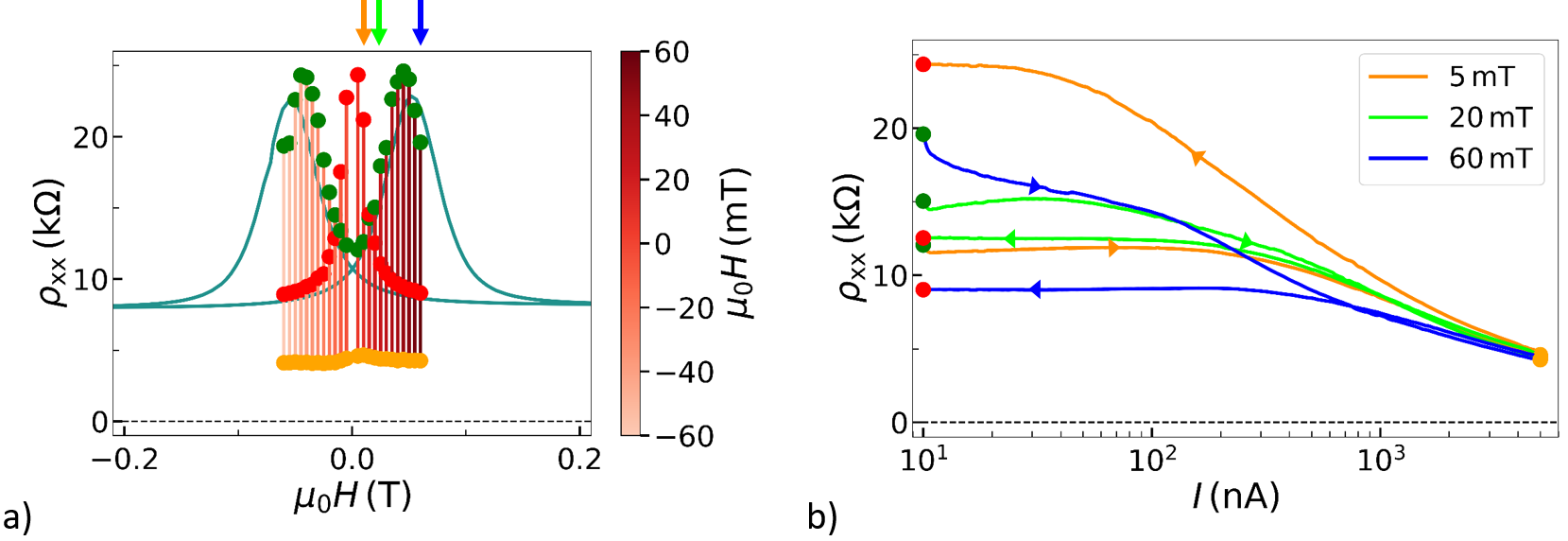}
\caption{Longitudinal current-induced magnetization switching experiment for different magnetic fields. a) The curves show current sweeps from $10\,$nA (green) to $5\,\upmu$A (yellow) back to $10\,$nA (red) at constant magnetic fields ranging from $-60\,$mT to $60\,$mT. The $10\,$nA hysteresis from Fig.~2~a) in the main text is plotted as a reference. b) The current dependence of three representative curves from a) are compared.}\label{Fig_Sandwich_CIMS_exp_xx}
\end{figure*}

In the main article the magnetization switching is discussed with a focus on the transversal signal, as there easily a switching of the sign is observed. In Figure~\ref{Fig_Sandwich_CIMS_exp_xx} the corresponding longitudinal data is shown, similar to Fig.~3 which shows the transversal data in the main text.

In Figure~\ref{Fig_Sandwich_CIMS_exp_xx} a) current sweeps at magnetic fields of $|\mu_0 H|\leq 60\,$mT with a stepping of $5\,$mT are shown and the $10\,$nA hysteresis from Fig.~2~a) in the main text is plotted as a reference. Colored dots mark the start at $10\,$nA (green), the maximum current position of $5\,\upmu$A (yellow), and the end at $10\,$nA (red). Except from the curves around $0\,$T the green and red points reproduce a similar behaviour as the reference curve. Thus, also in the longitudinal data the branch is switched only by applying a current for fixed magnetic fields.\\
In Figure~\ref{Fig_Sandwich_CIMS_exp_xx} b) three representative curves are plotted against the current. Interestingly, the resistivity is lowered for all curves when increasing the current. Although a decrease in longitudinal resistance could indicate an approaching of the QAHE, here, it is not the case, as the transversal resistivity decreases simultaneously~\cite{chang2013experimental}. For the curve taken at small external magnetic fields an increased resistivity is observed at the end of the sweep which indicates a vanishing chiral edge mode. Thus, this is consistent with the observations in the main article where for the transversal signal no realignment of the internal magnetic moments is concluded due to a too weak external magnetic field.
\noindent 

%

\end{document}